# A study on information behavior of scholars for article keywords selection

Zhixuan Lian

Business School of Nankai University, Tianjin, China 300071

Abstract

This project takes the factors of keyword selection behavior as the research object. Qualitative analysis methods such as interview and grounded theory were used to construct causal influence path model. Combined with computer simulation technology such as multi-agent simulation experiment method was used to study the factors of keyword selection from two dimensions of individual to group. The research was carried out according to the path of "factor analysis at individual level → macro situation simulation → optimization of scientific research data management". Based on the aforementioned review of existing researches and explanations of keywords selection, this study adopts a qualitative research design to expand the explanation, and macro simulation based on the results of qualitative research. There are two steps in this study, one is do interview with authors and then design macro simulation according the deductive and qualitative content analysis results.

## 1 introduction

Keywords in academic papers are in general considered as a core element that summarizes and represents the scientific publications' content, which are selected by the author of the paper. As these definitions imply, the practice of keywords selection can be thought of as a kind of democratic folksonomy tagging behavior with collective intelligence, because it builds a "folksonomy" from the bottom up and give better indexes to organize information, which bypasses the typical norms and practices of article indexing developed by LIS. Therefore, the theory of folksonomy tagging provides an opportunity to investigate user behavior in keywords selection.

In addition to employing keywords as tags of content, authors also use keywords as indexing concepts to organize academic resources during tagging process. Which means the recognized authority in academic can share common paradigm with authors by providing appropriate keywords to facilitate information access and resource discovery for authors. Authors can also select keywords to express their recognition and self-classification of the authority. Therefore, keywords are seemed as a set of knowledge representations of an author's cognitive understanding of academic resources and their content and in practice the semantic gap between academic indexing concepts and papers contents is inevitable. If academic retrieve system can clearly understand authors' keywords selection behaviors and their implicit cognition, they may be able to bridge the semantic gap between authors and themselves. Therefore, when we design information management system, we should consider the behavior of keywords selection, not only from the perspective of system but also from the perspective of individual. It still needs a study to elicit and build up the keywords selection models of article indexing in the individual perspective and the disparity according to authors' and system' keywords selection.

This article aims to explore authors' behavior models of article indexing in keyword selection. Thus, based on the research results, we can improve the automatic indexing technique of our articles retrieve system, taking into account the author's own tagging intentions and understanding of the content.

Related literature

Keywords selection are related to a number of objective factors such as citation counts, multidisciplinary crossing research collaborations, choice of journal for publication and contents readability([1][2][3][4][5]). In addition to the main subject characteristics of the article, this review considered only factors that the characteristics of the author's own information organization behavior. The factors considered in this review were classified into three groups: (i) factors extracted from the text of an article; (ii) factors extracted from the purpose of the author's selection, (iii) other factors.

2.1 factors extracted from the text of an article

The factors extracted from the text of an article include the (1) the content of the paper; (2) the background of the topic; and (3) the previous experience and knowledge of the authors. These 3 factors correspond to content channel, the background channel, and the prior knowledge channel([6]), which may play dissimilar roles and lead to different distributions average percentages of author keywords appearing in titles, abstracts, and both titles and abstracts are 31%, 52.1%, and 56.7%.([7]) in addition, it can also mine the factors of keywords according to the structural functions of the article content(u, Huang, Bu, and Cheng, (2018)[8]). In summary, the patterns of keywords extracted from the text can provide new insights into the keywords selection to explore the intrinsic behavior characteristics of the author's application of keywords and the intention of the paper itself expressed through keywords. Thus, the behavior features were quantified and integrated into a model of automatic keyword extraction, which can significantly improve the accuracy of automatic keyword extraction (Meng et al. (2017)[9]).

From the aforementioned studies, it can be seen that many investigations about the features of keywords focused on quantity and location distributions in papers to figure out the purpose of authors' selection. However, these results are interpreted from the perspective of the individual authors. To meet the needs from the management of vast amounts knowledge, Individual-perspective feature analysis is far from sufficient. There are still gaps in the exploring of macroscopic behavior characteristics making full use of the above results of the aforementioned studies.

---

[1] Duffy, R. D.,
Jadidian, A., Webster, G. D., & Sandell, K. J. (2011). The research productivity of academic psychologists: Assessment, trends, and best practice
recommendations.
Scientometrics, 89(1), 207–227.

[2] Egghe, L. (2006). Theory and practise of the g-index. Scientometrics, 69(1), 131–152.

[3] Gazni, A. (2011). Are the abstracts of high impact articles more readable? Investigating the evidence from top research institutions in the world. Journal of
Information Science, 37(3), 273–281.

[4] The impact of author-selected keywords on citation counts

[6] A Study on Mental Models of Taggers and Experts forvArticle Indexing Based on Analysis of Keyword Usage

[7] How do authors select keywords? A preliminary study of Author keyword selection behavior

[8] Lu, W., Huang, Y., Bu, Y., & Cheng, Q. (2018). Functional structure identification of scientific documents in computer science. Scientometrics, 115(1),
463–486.

[9] Meng, R., Zhao, S., Han, S., He, D., Brusilovsky, P., & Chi, Y. (2017). Deep keyphrase generation. In Annual Meeting of the Association for Computational
Linguistics (pp. 582–592).

2.2 factors extracted from the behavior of the author's selection

As an important way of organizing information and knowledge, the keywords selection can be regarded as the tagging behavior of the author's own paper to represent the cognitive understanding of information resources and their content (Chen & Ke, 2014[10]). Examples of this view are always focus on the difference between authors with different cognitive levels or backgrounds(Tsai, Hwang, and Tang, (2011)[11]) and explore their mental models (Ke and Chen (2012)[12] Chen & Ke, 2014[13]).

At the same time, the proper selection of keywords is proved having a lot to do with the increase of citation counts ([14]). For increasing the citation counts the authors tend to promote the visibility of paper though selecting the popular keywords ([15]). and the main steam of this view is concentrated on predicting citations using keywords as a tool, including(i) the citation network structure of keywords ([16]), (ii)Keyword growth (i.e., the relative increase or decrease in the presence statistics of an underlying keyword over a given period of time); (iii) Keyword diversity (i.e., the level of variety in a set of author-selected keywords); (iv) Number of keywords; and percentage of new keywords([17]).

Because the collective behavior in keyword selection can reflect the fact that the science paradigm recognition by the research community, on account of the theory of scientific community. So that whether it's from a tagging behavior or from a citation increasing behavior, individual behavior will be considered at the macro level in this paper. Improving our automatic indexing technology from collective behavior factors will be more in line with the needs of mass knowledge management.

With reference to the research outlined earlier, this study proposes the following research question:

RQ1: what factors do authors consider when making indexing?

RQ2: Affecting by these factors, what is the law of the author's keyword selection behavior from a macro perspective?

---

[10] Chen, Y. N., & Ke, H. R. (2014). A Study on Mental Models of Taggers and Experts for Article Indexing Based on Analysis of Keyword Usage. Journal of the Association for Information Science and Technology, 65(8), 1675–1694.

[11] Tsai, L. C., Hwang, S. L., & Tang, K. H. (2011). Analysis of keyword-based tagging behaviors of experts and novices. Online Information Review, 35(2), 272–290.

[12] Ke, H. R., & Chen, Y. N. (2012). Structure and pattern of social tags for keyword selection behaviors. Scientometrics, 92(1), 43–62.

[13] Chen, Y. N., & Ke, H. R. (2014). A Study on Mental Models of Taggers and Experts for Article Indexing Based on Analysis of Keyword Usage. Journal of the Association for Information Science and Technology, 65(8), 1675–1694.

[14] The impact of author-selected keywords on citation counts

[15] Identification of highly-cited papers using topic-model-based and bibliometric features: the consideration of keyword popularity

[16] Predicting scientific research trends based on link prediction in keyword networks

[17] The impact of author-selected keywords on citation counts

2.3 The authoritative methods of Selecting keywords for domain analysis in bibliometrics
Identifying important publication keywords in bibliometrics

Methods
3.1 General research design
　　Based on the aforementioned review of existing researches and explanations of keywords selection, this study adopts a qualitative research design to expand the explanation.
3.2 Participants and data collection
　　Having decided upon the general methodological approach, this study invited 15 authors of different fields of science, gender and faculty positions to provide detail descriptions of their keywords selection at writing， submitting and modifying. The sample was formed through a combination of convenience sampling and snowballing. A profile of the resulting sample is shown in Table I(including 7 professors and 8 PHD students covering 13 scientific fields). Data were collected through semi-structured interview, for example, the interviewees were asked to describe their experience of keyword selection after the paper completion, and explore their view on the role of keywords, the characteristics of the selected keywords, and whether to quote a new keyword or not and so on. For each type of experience, they were asked to describe the details of the backgrounds of the papers, the aims of keywords to express and other factors they considering. The interviews took from 20min to 1h, with the majority taking around 20min with recording upon the participants' consent.

　　Following the general qualitative data analysis approach and grounded theory principles, the interview data were analyzed through multiple rounds of coding to supplement the previous studies. The initial coding followed no theoretical guidance other than the research aim itself, which is to develop a theoretical framework that integrates the structural and agentic dimensions to explain keywords selection related factors. Next, considering the previous studies, t the results of previous studies will be verified, and the parts that have not been studied will be explored and put forward the causal hypothesis to figure out the proposed RQs.

Table I List of interviewees

| Code | Record of formal schooling | position | Science field |
|---|---|---|---|
| 1 | doctor | professor | management |
| 2 | doctor | engineer | engineering |
| 3 | doctor | Research assistant | economics |
| 4 | doctor | The editor | engineering |
| 5 | doctor | doctor | medicine |
| 6 | doctor | professor | mathematics |
| 7 | doctor | professor | Philosophy |

4. data analysis from the interview and hypothesize

Data analysis at this stage first broke each participant's interview accounts into descriptions of their keywords selection experience stages. it then assigned a code to everything that the participant mentioned so long as it fell into the following categories: (1) external condition factors affecting the previous experience and knowledge of the authors (2)the internal driving factors compelling or refraining the participant's selecting a particular keyword (3) keywords selection actions process themselves.

During the second round of coding, these free codes were compared with each other and were merged to from broader codes. The resulting codes and the data behind them were further compared to ascertain the properties, range, and variations of what they describe. During the third round of coding, the original interview transcripts were examined again. This time focusing mainly on the relationships among the obtained factors and try to explain them by folksonomy tagging theory. This round was divided into two steps. First, new codes were developed to describe these relationships; second, the resulting codes and the original data behind them were reexamined to identify conditions, contexts and qualifications of each relationship. By the end of this round, 3 major categories were established and their relationships were identified, resource, users and tags.

During the fourth round of coding, the above categories and their interrelationships were further examined to see how they were integrated to form explanations about the folksonomy tagging for keywords selection. Three major themes of factors emerged at this stage, which formed the backbone of the integrative framework proposed in this paper:The following sections explicate these themes in detail.

5 Emergence of the key concepts from the data

5.1concepts of keywords selection explained by folksonomy tagging theory

As mentioned in the methodology section, data analysis towards an integrative framework for keywords selection explained by folksonomy tagging theory started from an open coding of interviewees' accounts of their keyword selection at different stages of using academic resources. It was obvious from the beginning that keywords selection is a dynamic process of repeated adjustment. Authors often use keywords during literature research which put a more or less influence on them when they formulate their own keywords. Examples of such acts include searching the key words in the science field of the paper, searching the key words of the journal which is as a target to submit the paper and use these key words for literature retrieval. These were coded as keywords selection during data analysis. The aggregation of a person's keywords selection in a certain period or field was then called his/her information behavior for that period field. Examples of an individual's information behavior include his/her academic resource reviewing activity, tagging activity according to the paper content, selection and modification activity for submitting the paper, etc.

After instructions of what is keywords selection in this paper, it will be put in a folksonomy tagging perspective. As folksonomy tagging theories indicate and as the above data analysis shows, the folksonomy tagging theory consists of three information access channels: resource, tags and users. When initial codes for objective factors affecting a person's keywords selection were examined, it became fairly obvious that most objective factors affecting selection are related either to selection practices themselves or to the aim of authors or the influence of the mainstream view of academic, which are corresponding to three channels.

At first, the concepts of resource of folksonomy emerged from Various types of information stored in a collaborative tagging system, which has a typical feature that sharing from and managed by users. The academic resource retrieving is not only an understanding of the information tagged by others, but also a preset understanding of their own information, for instance that almost all interviewees mentioned that she or he always searches papers in the same direction according to her research direction when finishing her own papers. The keywords used in retrieval are very related to the keywords tagged by oneself. From a macro perspective, the relevant literature in a scientific field, both published and under research, can be regarded as the concept of 'resources', which can explain the interviewees' descriptions of their Academic Resources retrieving at different stages.

The concept of 'user' refers to the object of a system service, including the creator, annotator, or consumer of a resource. There are often several social groups in the collaborative tagging system, each of which is composed of users with the same interest field or knowledge structure. The group members can connect their information space by a forum platform for speech and communication. Academic conferences and journals often serve as platforms for such exchanges. Almost all interviewees mentioned that the purpose of finishing a paper is to publish and communicate with their peers, and they all took the attitudes of readers in the same field into consideration when selecting keywords. Therefore, as the expounds above, the keyword selection and accepted is individuals who act and, hence, generate actions that aggregate into activities and organized activities of a group of academic community. This study terms these arrangements promoting the keywords accepted as the User-related factors.

'tag' is the bridge connecting the 'users' and 'resource'. It is the identification given by users to refers to a specific resource and sharing in collaborative tagging system. In a collaborative tagging system, tags can aggregate similar resources and contribute to the formation of a science community. The behavior of a large number of users adding tags to a large number of resources forms sociality through collision and integration, and information resources can be automatically classified under the condition of tags. One experience from interviewee 1 provided the cases in point. She recalled an experience of research in the field of e-government when there were few scholars in the field. As a pioneer in e-government research, she and her team have published a number of papers that have also attracted interest from scholars in the field, More and more academic research on the keywords related to 'e-government' has promoted the formation of a scientific community in this field. At the same time, she gives us an example to the contrary. Some of her papers were not retrieved because she did not cite "e-government" as a keyword. This study terms the factors related to 'tag' as a channel of interaction between users and resources by observing the formation of scientific community and the integration of resources.

5.2 the factors about 'resource'

This study examined the information in the code that involves 'resources' firstly. The 'background of the topic' has been adequate studied. As well as we have an important hypothesis that is the influence of interdisciplinary science fields background based on previous studies. For instances that the interviewee 2,3,4 are all interdisciplinary scholars. The interviewee 2 gave an experience that he attempted to use machine learning methods to process huge amounts of data from epidemiological surveys. His research topic needs to carry out literature retrieval from two aspects of research object and method which are not in the same science field category. After taking this into account, the final accepted keywords addressed both two science fields. Similarly, the research area of interviewee 3 is the intersection of environmental science and economics, and the research area of interviewee 4 is the intersection of communication science and information science. Their selection of keywords also reflected the review of resources in two science field. To further understand how they are affected by 'resources', we used a snowballing interview to know deeper. interviewee 4 recalled that must be unequivocal what is the hot question in the field of her research object, and whether her paper can answer or supplement this question well. And the related keywords existed in the 'resource' would be give priority to adopt for creating a new tag can be risky not being accepted. And then the innovative of method would be considered especially in examining that whether the method has been accepted before. Of course, applications of recently invented methods are always more popular. Interviewee 2 added a detail that he always browsed the journals about CV (Computer Vision) to find inspiration for solving the epidemic investigation. At first the intersecting research in multiple fields have also more varied keywords combination. So the more keywords that can be selected in interdisciplinary research, the more difficult it is to reach a consensus on a particular keyword in that field. then, authors tend to accept popular keywords which can be to deduce that the behavior of keyword selection is influenced by the whole resource environment. From the above we can make a conclusion.Multiple academic resources will bring a variety of popular keywords into academic community, and promote the diversification of keyword selection behavior.

5.3the factors about 'user'
   As shown in the previous section, when talking about their experience about particular keywords selection, the all interviewees refer time and again to other scholars' role as the potential audiences. Further examination shows that when they do the literature retrieving and keyword tagging, they take into account who will use their keywords during retrieving, reviewing and citing. But the same attempt to attract the attention of other scholars has led to two different kinds of behavior, the one is that some scholars tend to cite a key word that is not in their field to attract attention. As an instance that the interviewee 7 came up with some new key words that encapsulate his own ideas when he researched current political events. The interviewee 1 and 4 had the similar experience with introducing or creating a new keyword. as for other interviewees, the popular vocabulary recognized by the scientific community is generally adopted. Because different condition embedded behaviors of keywords selection to different degrees and in different ways. Interviewee 1, 4 and 7 Are heavyweight scholars in their fields in hence that they have enough traction and followers in their own right to reinforce their views. Propose new ideas to broaden the field of vision and put forward new research framework can improve the retrieval rate and citation rate more. Young scholars are often working with existing frameworks and using popular keywords to improve their retrieval rate. From the above we can make a conclusion. the keywords selection of heavyweight scholars can affect the academic environment to some extent, while young scholars are often affected by the environment.

5.4 the factors about 'tag'

Not just the status of the scholars themselves, their previous experience and knowledge are also the factors that affect the keywords selection which are defined as the channel of authors previous experience by the previous studies18. And then the references of papers can be used to represent the prior knowledge channel for references represent intellectual or cognitive influence on the scientific work, which can also indicate the previous experience and knowledge of authors concerning the scientific work as Bornmann and Daniel (2008) argued that [19]. The interviewee7 provided a view that the collective behavior in keyword selection can reflect the role of previous experience factors of a science field in the process of scientific understanding development. His research is focused on reviewing and developing disciplinary theories, with emphasis on referring to concepts mentioned in other literature in the same field when he chooses his keywords. When he develops a new concept, he must also ensure that his own concept is supported by previous research within the framework of existing research. Therefore, the keywords he accepted can be seen as a summary and development of previous studies he cited.

According to the theory of scientific community, these members of a scientific community always use a common paradigm, because it is pointed out that a paradigm is only something shared by members. The keywords can be regarded as something shared by the science community. Therefore, by observing and simulating the consistency of scholars' behavior in keyword selection on a macro perspective, it is can be predicted that the formation and development of the scientific community, so as to clarify the development process of some a branch of science paradigm, which is playing the role of the 'tag' as a channel for the formation of user groups based on the insights in common.

The interviewee 7 also noted that, the acceptance of the key word is often not very good when he comes up with a key word about a new concept However, after the paper was published for a period of time, it had a certain number of citations, and the usage rate of this keyword became higher. And there is a certain correlation between the reception of key words and the citation of his article which is called branching from his interview. from his experience, it is can be observed that an innovation of a paradigm can be regarded as the formation process of a new scientific community, which diffusion in spreads among members of a society over time through a specific channel20. There was very little paradigm-innovation acceptation in the early stages of diffusion before the pioneers in scientific community who adopt the paradigm-innovation. As diffusion proceeds, the second tier of individuals associated with pioneers gradually adopt the new paradigm and the diffusion is continuing layer upon layer21. From the above we can make a conclusion：The keywords selection is related to the structure of the citation network.

---

[18] How do authors select keywords? A preliminary study of Author keyword selection behavior
[19] A quantitative exploration on reasons for citing articles from the perspective of cited authors
[20] Rogers E M. Diffusion of Innovations[M]. New York: Free Press, 1983.
[21] Banerjee A V. A Simple Model of Herd Behavior[ J]. Quarterly Journal of Economics, 1992, 107(3):797 - 817.